\documentclass[aps,prl,twocolumn,superscriptaddress,showpacs]{revtex4}
\usepackage{graphicx,psfrag}
\DeclareGraphicsExtensions{.ps, .eps}

\begin{document}

\title{Muon diffusion and electronic magnetism in Y$_2$Ti$_2$O$_7$}

\author{J.~A.~Rodriguez}
\email{jose.rodriguez@psi.ch}
\affiliation{Laboratory for Muon-Spin Spectroscopy, Paul Scherrer Institut, CH-5232 Villigen-PSI, Switzerland}
\author {A.~Yaouanc}
\affiliation{Laboratory for Muon-Spin Spectroscopy, Paul Scherrer Institut, CH-5232 Villigen-PSI, Switzerland}
\affiliation{Institut Nanosciences et Cryog\'enie, SPSMS, CEA and University Joseph Fourier, F-38054 Grenoble, France}
\author{B.~Barbara}
\affiliation{Institut N\'eel, CNRS and universit\'e Joseph Fourier, B.P. 166, 38042 Grenoble, Cedex 09, France}
\author {E.~Pomjakushina}
\affiliation{Laboratory for Developments and Methods, Paul Scherrer Institut, CH-5232 Villigen PSI, Switzerland}
\author{P.Qu\'emerais}
\affiliation{Institut N\'eel, CNRS and universit\'e Joseph Fourier, B.P. 166, 38042 Grenoble, Cedex 09, France}
\author{Z.~Salman}
\affiliation{Laboratory for Muon-Spin Spectroscopy, Paul Scherrer Institut, CH-5232 Villigen-PSI, Switzerland}

\date{\today}

\begin{abstract}
  We report a $\mu$SR study in a Y$_2$Ti$_2$O$_7$ single crystal. We
  observe slow local field fluctuations at low temperature which
  become faster as the temperature is increased. Our analysis suggests
  that muon diffusion is present in this system and becomes small
  below 40~K and therefore incoherent. A surprisingly strong
  electronic magnetic signal is observed with features typical for
  muons thermally diffusing towards magnetic traps below $\approx
  100$~K and released from them above this temperature.  We attribute
  the traps to Ti$^{3+}$ defects in the diluted limit. Our
  observations are highly relevant to the persistent spin dynamics
  debate on $R_2$Ti$_2$O$_7$ pyrochlores and their crystal quality.
\end{abstract}
\pacs{76.75.+i, 75.10.Jm, 75.40.Gb}
\maketitle

The rare-earth titanates and stanates series of compounds,
$R_2M_2$O$_7$ ($R$ is a rare-earth ion and $M$ is Ti or Sn), which
crystallize in the pyrochlore crystal structure (space group $Fd{\bar
  3}m$), are prone to strong geometrical frustration.\cite{Gardner10}
Their study has revealed a wealth of exotic magnetic properties. These
include (i) the spin-ice ground state of Ho$_2$Ti$_2$O$_7$ and
Dy$_2$Ti$_2$O$_7$,\cite{Harris97,Ramirez99,Dunsiger2011} (ii) the
ground state reached by Yb$_2$Ti$_2$O$_7$ after a sharp transition in
the spin dynamics finger-printed by a pronounced peak in the specific
heat,\cite{Hodges02} (iii) the unconventional dynamical ground state
of Tb$_2$Sn$_2$O$_7$ for which magnetic Bragg reflections are observed
by neutron diffraction,\cite{Mirebeau05} while no spontaneous magnetic
field is measured by the zero field (ZF) muon spin relaxation
($\mu$SR) technique,\cite{Dalmas06} (iv) the persistent spin dynamics
(PSD) detected in the ordered states of Gd$_2$Sn$_2$O$_7$,
Gd$_2$Ti$_2$O$_7$ and
Er$_2$Ti$_2$O$_7$,\cite{Bertin02,Yaouanc05a,Chapuis09b,Lago05,Dalmas12}
and the non-ordered PSD state of
Tb$_2$Ti$_2$O$_7$.\cite{Yaouanc11a,Gardner99}

The first report of PSD was for SrCr$_8$Ga$_4$O$_{19}$ using
$\mu$SR.\cite{Uemura94} It was found that this kagome compound does
not exhibit magnetic Bragg reflections down to 50~mK. Although mostly
observed by $\mu$SR, PSD has been also proposed by other techniques,
e.g., in Gd$_2$Sn$_2$O$_7$ using $^{155}$Gd M\"ossbauer
spectroscopy.\cite{Bertin02} Both $\mu$SR and the neutron spin echo
have observed PSD in Tb$_2$Sn$_2$O$_7$\cite{Dalmas06,Rule09b} as well
as Dy$_2$Ti$_2$O$_7$.\cite{Lago07,Dunsiger2011,Gardner11} However, the
existence and nature of PSD is still under
debate\cite{Yaouanc05a,McClarty11}.  Recently, it was suggested that
PSD does not exist for Dy$_2$Ti$_2$O$_7$ and that the observed
relaxation results from coherent muon diffusion.\cite{Quemerais12} In
that work it was proposed to perform $\mu$SR measurements in
Y$_2$Ti$_2$O$_7$ to investigate the possibility of coherent muon
diffusion in a compound where frustrated magnetism does not play any
role as no 4f-magnetism should be present. In this letter we report ZF
and Longitudinal Field (LF) $\mu$SR measurements on Y$_2$Ti$_2$O$_7$.
We show that there is no detectable neutral muonium formation, and
that the positive-muon hopping rate at low temperature is much smaller
than what was proposed for Dy$_2$Ti$_2$O$_7$.\cite{Quemerais12} Also,
we have observed a relatively strong muon spin relaxation which we
attribute to a small density of Ti$^{3+}$ magnetic defects. Note that
this system has already been studied by Dunsiger using
$\mu$SR.\cite{Dunsiger00a} Nevertheless, that study was less extensive
than the one reported here and did not consider the possibility of
coherent diffusion at low temperature. When a comparison is possible,
our data and that from Dunsiger are similar.

\begin{figure}[htb]
\includegraphics[width=0.88\columnwidth]{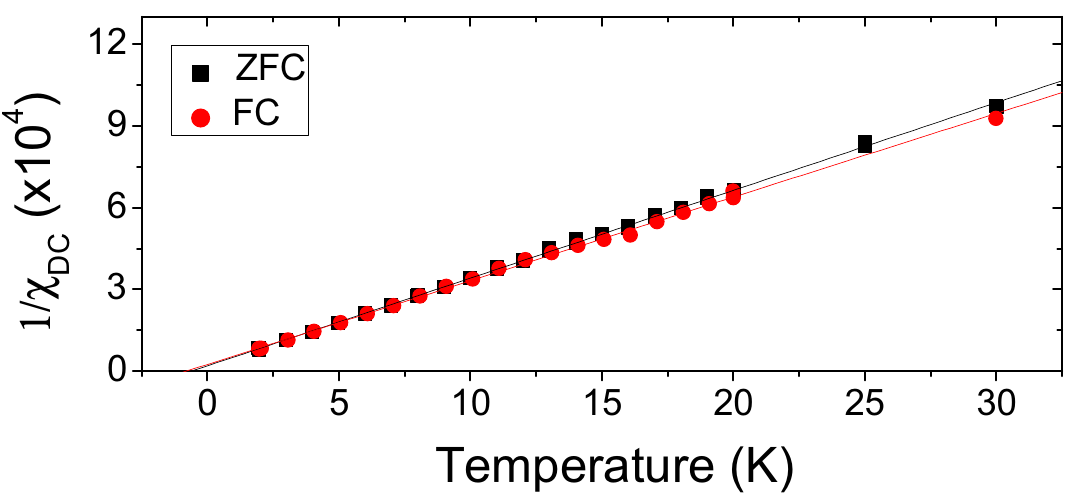}
\caption{Inverse susceptibility in SI units as a function of
  temperature in zero field cooled (ZFC) and field cooled (FC)
  protocols. It was verified that the applied field was small enough
  to be in the linear regime. We fit the data to
  $\chi_{\mathrm{DC}}=C_{\mathrm{CW}}/(T-T_{\mathrm{CW}})$ and obtain
  $C_{\mathrm{CW}}$= 2.45(5)$\times$10$^{-4}$~K and
  $T_{\mathrm{CW}}$=-0.59(2)~K. Assuming that the magnetic moments in
  the system are equal to 1$\mu_\mathrm{B}$, we can estimate their
  concentration at the yttrium site (see discussion) as $y=3 \:
  C_{\mathrm{CW}} \: a^3 \, k_\mathrm{B} / 16 \: \mu_0 \:
  \mu_\mathrm{B}^2$, where $a$ is the lattice constant. This formula
  gives $y$=0.60(1)\%.}
\label{Y2Ti2O7_magnetization}
\end{figure}

Polycrystalline Y$_2$Ti$_2$O$_7$ was prepared by a solid state
reaction. Starting materials of Y$_2$O$_3$ and TiO$_2$ with 99.99\%
purity were mixed and ground. They were then heat treated at
900-1150$^\circ$C in air for more than 100h with several intermediate
grindings. The resulting powder was hydrostatically pressed in the
form of rods (8 mm in diameter and 60 mm in length). The rods were
subsequently sintered at 1150$^\circ$C during 15h. The crystal growth
was done using an optical floating zone furnace with four 1000W
halogen lamps as a heat source. The growing conditions were: growth
rate of 10~mm/h, feeding and seeding rods were rotated at about 20~rpm
in opposite directions (to have homogeneity of the liquid), growth
done in 4~bar pressure of an argon and oxygen mixture (50:50). The
crystal was post-annealed at 1150$^\circ$C in argon for 15h in order
to remove possible over stoichiometric oxygen. Phase purity of the
grown crystal was checked with conventional powder x-ray
diffractometer and the obtained lattice parameter $a$=10.099\AA ~is in
good agreement with the literature.\cite{Subramanian83} The crystal
was then aligned using an X-ray Laue camera, and magnetization
measurements were performed
down to 1.9~K (see Fig.~\ref{Y2Ti2O7_magnetization}).

The $\mu$SR measurements were carried out on the General Purpose
Spectrometer (GPS) at the Swiss Muon Source facility of the Paul
Scherrer Institut (Switzerland). Most of the measurements were done in
the LF geometry for which the initial muon spin and the external
magnetic field ${\bf B}_{\rm ext}$ are
parallel.\cite{Schenck95,Yaouanc11} We define the $Z$ axis to be
parallel to the initial muon spin direction, with the positron
detectors centered along this axis. With this geometry the measured
asymmetry ({\it i.e.} $\mu$SR signal) is written as $a_0P_Z(t)$, where
$a_0$ is the initial asymmetry (a characteristic of the spectrometer
and the geometry of the sample), and $P_Z(t)$ is the longitudinal muon
polarization function which contains information on the local magnetic
fields in sample. In addition, three measurements were done with the
Transverse Field (TF) geometry where ${\bf B}_{\rm ext}$ is
perpendicular to $Z$.

\begin{figure}[htb]
\includegraphics[width=0.88\columnwidth]{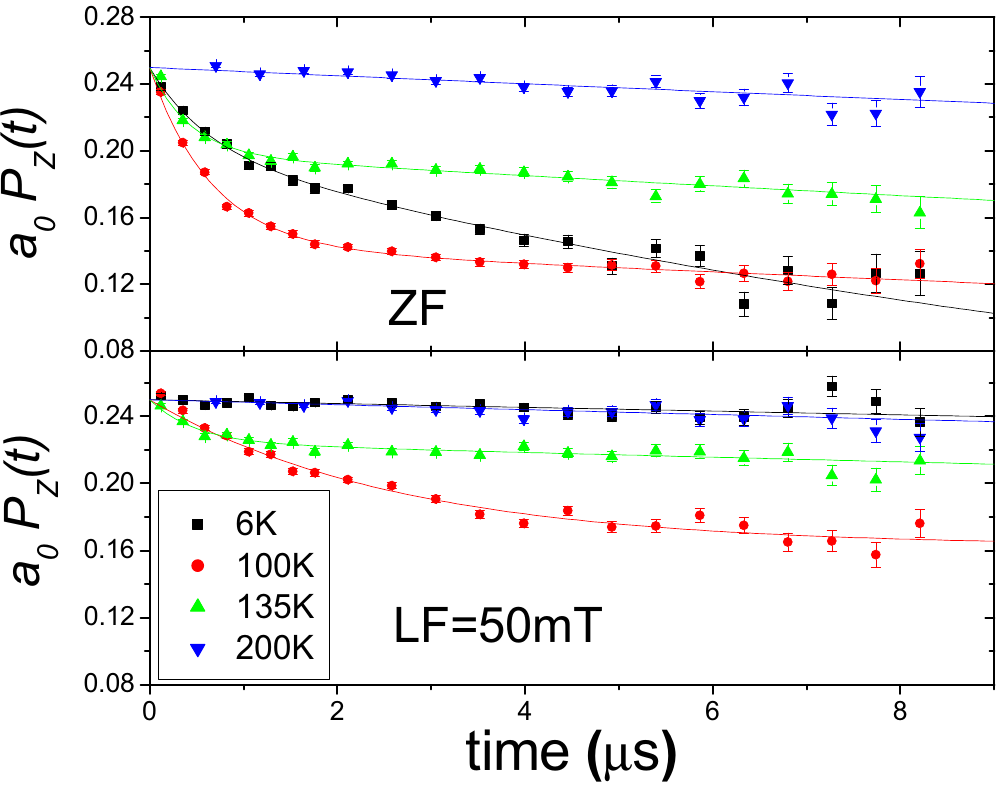}
\caption{(Color on-line) ZF (upper panel) and LF=50mT (lower panel)
  spectra as a function of temperature. The initial muon beam
  polarization is parallel to the $[100]$ crystal direction. The lines
  are fits as described in the text.}
\label{Y2Ti2O7_ZF_LF50_spectra}
\end{figure}

\begin{figure}[htb]
\includegraphics[width=0.88\columnwidth]{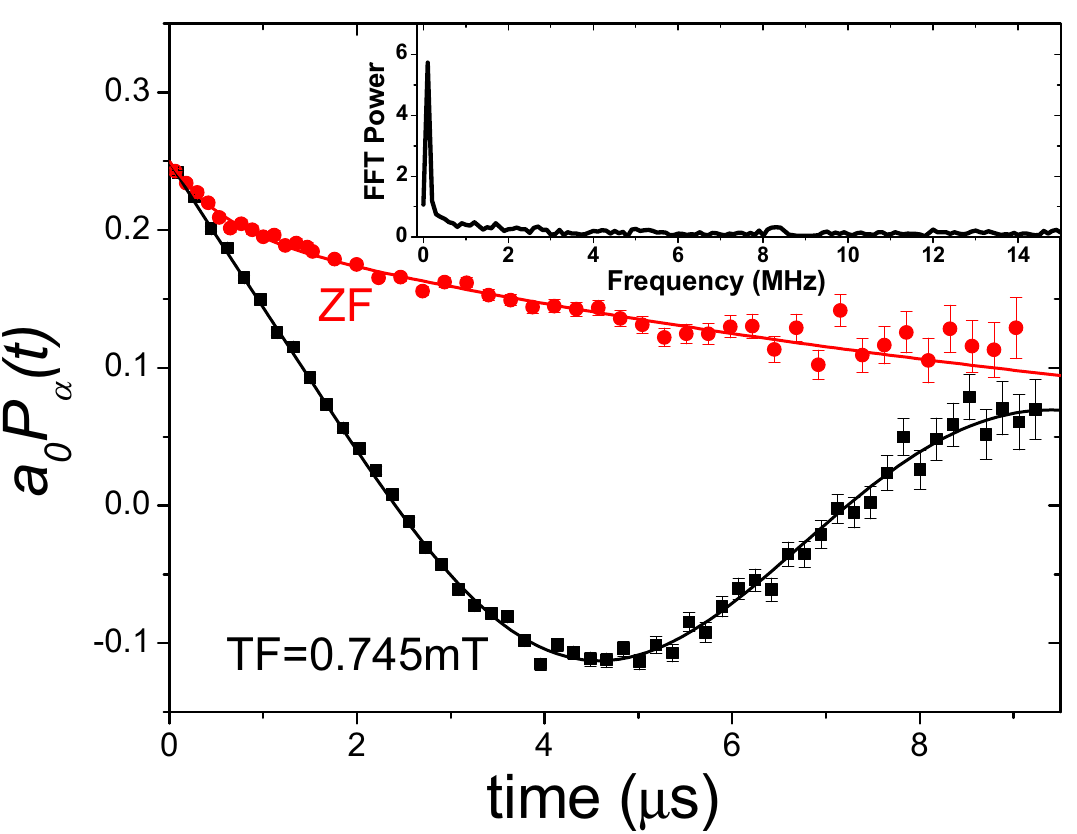}
\caption{(Color on-line) TF ($0.745$~mT) and ZF spectra measured at
  2.4~K. The inset shows the Fourier transform of the TF spectrum,
  where a peak is observed at 0.101~MHz corresponding to the applied
  TF.}
\label{Y2Ti2O7_TF_spectra}
\end{figure}

It is important to know if there is muonium formation in
Y$_2$Ti$_2$O$_7$, as this entity is known to diffuse coherently at low
temperature in some systems.\cite{Yaouanc11} Typical ZF and LF (50~mT)
spectra measured at different temperatures are shown in
Fig.\ref{Y2Ti2O7_ZF_LF50_spectra}. The initial asymmetry is
temperature and field independent with a value $a_0 \approx 0.25$
typical for the GPS. This is a strong indication that there is no
muonium formation in the whole temperature range. As expected for the
absence of muonium, no re-polarization (a recovery of the missing
asymmetry) by the LF is observed. The absence of muonium is further
supported by the TF and ZF signals at 2.4~K
(Fig.~\ref{Y2Ti2O7_TF_spectra}), where the initial asymmetry is found
to be the same for the two spectra and the only frequency observed in
TF is that of the applied field. This is further supported by Fourier
Transform of the TF signal shown in the inset of
Fig.~\ref{Y2Ti2O7_TF_spectra}. If an appreciable vacuum-muonium
fraction was present, we would expect to observe an oscillation at a
frequency $ \approx 103 \gamma_\mu B_{\rm ext}/(2 \pi) = 9.8$~MHz
($\gamma_\mu =851.615 \, {\rm Mrad} \, {\rm s}^{-1} \, {\rm T}^{-1}$)
in the Fourier spectrum,\cite{Yaouanc11} and/or a reduction of the
initial asymmetry due to the fast muonium precession. Therefore, we
find no evidence of neutral muonium states in Y$_2$Ti$_2$O$_7$. The
reason for the muonium absence in the pyrochlore oxides is not
understood, and it is certainly a subject of much interest.

\begin{figure}
\includegraphics[width=0.88\columnwidth]{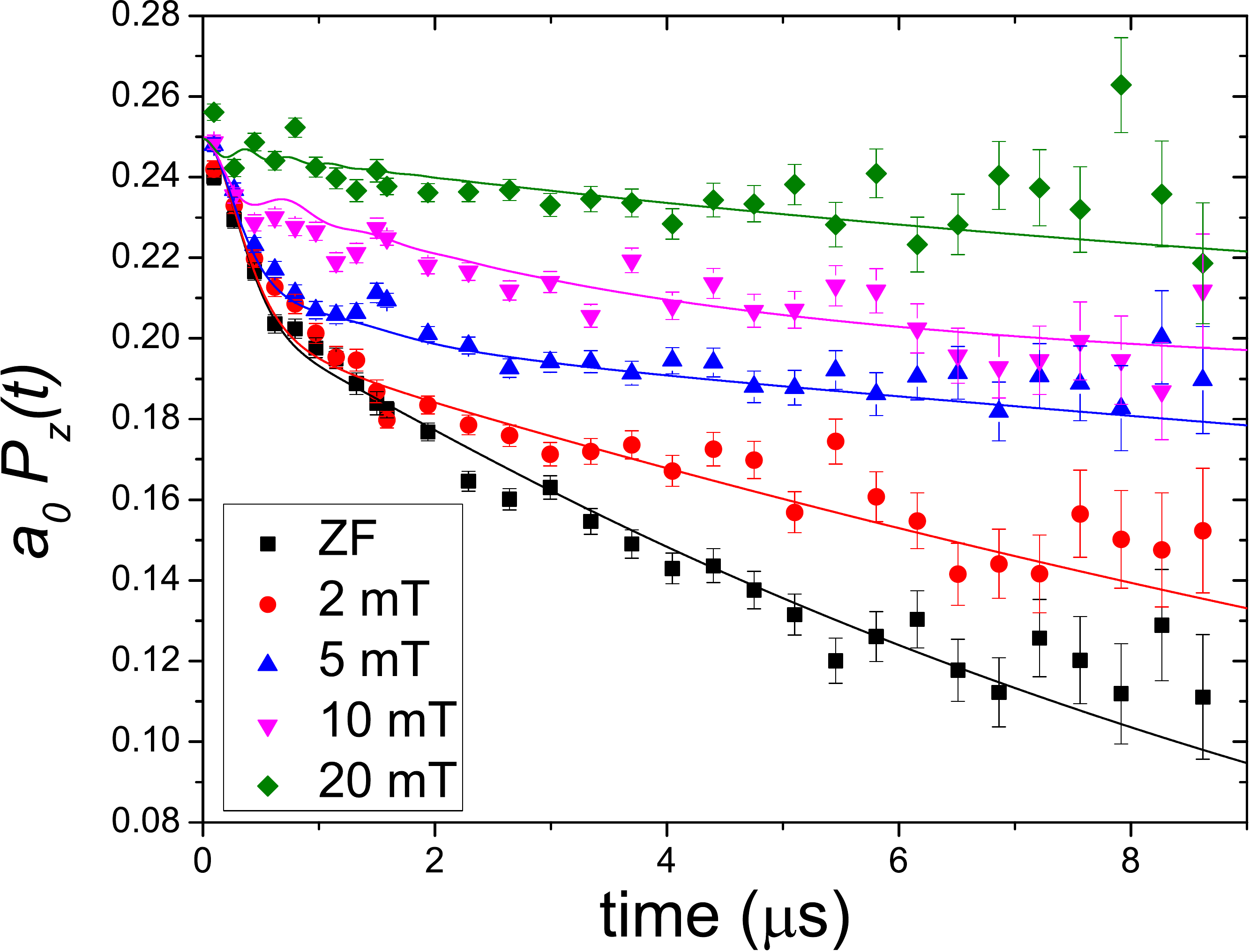}
\caption{(Color on-line) The field dependence of the LF data at 2.4K.
  The solid lies are fits as described in the text.}
\label{Y2Ti2O7_LF_spectra_decoupling}
\end{figure}

We turn now to discussing the ZF and LF data in
Fig.~\ref{Y2Ti2O7_ZF_LF50_spectra}, where we observe a surprisingly
high relaxation rate. {\it A priori}, electronic magnetism should not
be present in Y$_2$Ti$_2$O$_7$ since both Y$^{3+}$ and Ti$^{4+}$ ions
are non-magnetic.
Therefore, only nuclear magnetic moments should be present in this
system. However, as shown in Fig.\ref{Y2Ti2O7_magnetization}, the
susceptibility shows a relatively strong paramagnetism. Also, as shown
in Fig.~\ref{Y2Ti2O7_LF_spectra_decoupling}, not even a field $B_{\rm
  ext} = 20$~mT can completely decouple the relaxation of the $\mu$SR
signal (usually less than 3~mT are needed to decouple nuclear
fields,\cite{Karlsson95} which we estimated to be $\approx20$~$\mu$T
in Y$_2$Ti$_2$O$_7$). Hence, we conclude that the relaxation is of
electronic origin as proposed by Dunsiger.\cite{Dunsiger00a}

To understand the origin of the electronic magnetic moments we first
discuss the Y$_2$Ti$_2$O$_7$ crystal structure.  In general, we note
that the ternary pyrochlore compounds are of the general formula
$A_2B_2$O$_7$.\cite{Subramanian83} The $B$ element can be a transition
metal with a variable oxidation state. This gives the possibility of
substitution on the $B$ site (see Ref.~[\onlinecite{Ueland10}] and
references therein). The structure also tolerates vacancies at the $A$
and O sites to a certain extent. Around the $A$ and $B$ cations one
finds 8- and 6-fold coordination polyhedron of oxygen, respectively.
Focusing on Y$_2$Ti$_2$O$_7$, in terms of the oxidation states we have
Y$_2^{3+}$Ti$_2^{4+}$O$_7^{2+}$.  Ti$^{4+}$ is diamagnetic since it is
in a $d^0$ configuration. Depending on its coordination the Ti$^{4+}$
ionic radius is 74.4 or 88~pm, which is smaller than the Y$^{3+}$
ionic radius of 104 or 115.9~pm. Now we consider a magnetic
configuration for at least one of the three involved elements. The
obvious candidate is Ti$^{3+}$ which has about the same ionic radius
as Ti$^{4+}$. This ion is known to be rather unstable\cite{Abragam70}
and it has been suggested to locate it in the $A$
site.\cite{Subramanian83} In this case we can write (Y$_{2 -
  x}$Ti$_x$)Ti$_2$O$_7$, where we have neglected the substitution of
Y$^{3+}$ in the $B$. This does not change our conclusions since
Y$^{3+}$ is non-magnetic.\cite{Ross12,Revell13} Ti$^{3+}$ has an
electronic configuration $d^1$ and therefore, as a free ion, carries
one Bohr magneton. Another possibility, which is less likely though
cannot be rule out, is O vacancies in the structure. Such defects will
also introduce two Ti$^{3+}$ magnetic moments for each
vacancy.\cite{Ross12} Note that both types of defects will produce
essentially the same effect on the implanted muons and cannot be
distinguished in our measurements.

We now consider qualitatively the ZF and 50~mT LF spectra of
Fig.~\ref{Y2Ti2O7_ZF_LF50_spectra}. In ZF and low temperatures we
observe a signal with a fast and a slow relaxing components. As the
temperature is increased, the relaxation rate of the fast component
decreases slightly and then increases until this component disappears
above 150~K. The relaxation of the slow component decreases
continuously with temperature. In the LF measurements the overall
relaxation rate start increasing as the temperature is increased,
peaks at 100~K, and then it decreases to a small value above 150~K.
Referring to the original work of Borghini {\it et
  al.}\cite{Borghini78} and the latter works reviewed in
Ref.~[\onlinecite{Karlsson95}], the temperature dependence of the
spectra is typical for a muon trapping/detrapping effect.  At low
temperature the overall relaxation is small because a big fraction of
muons are implanted ``far'' from Ti$^{3+}$ magnetic defects.  As the
temperature is increased, the relaxation increases because muons
diffuse towards magnetic defects and become trapped. At even higher
temperatures, trapping/detrapping become faster leading to a decrease
of the relaxation rate due to the fast fluctuating magnetic field
sensed by the muons. The 50~mT field is sufficient to quench the
relaxation except around 100~K, where a $B_{\rm ext} = 200$~mT was
needed to suppress the relaxation at this temperature. This is
consistent with muons diffusing and accumulating near magnetic defects
at this temperature, {\it i.e.} the trapping rate is higher than the
detrapping rate.

\begin{figure}
\includegraphics[width=0.88\columnwidth]{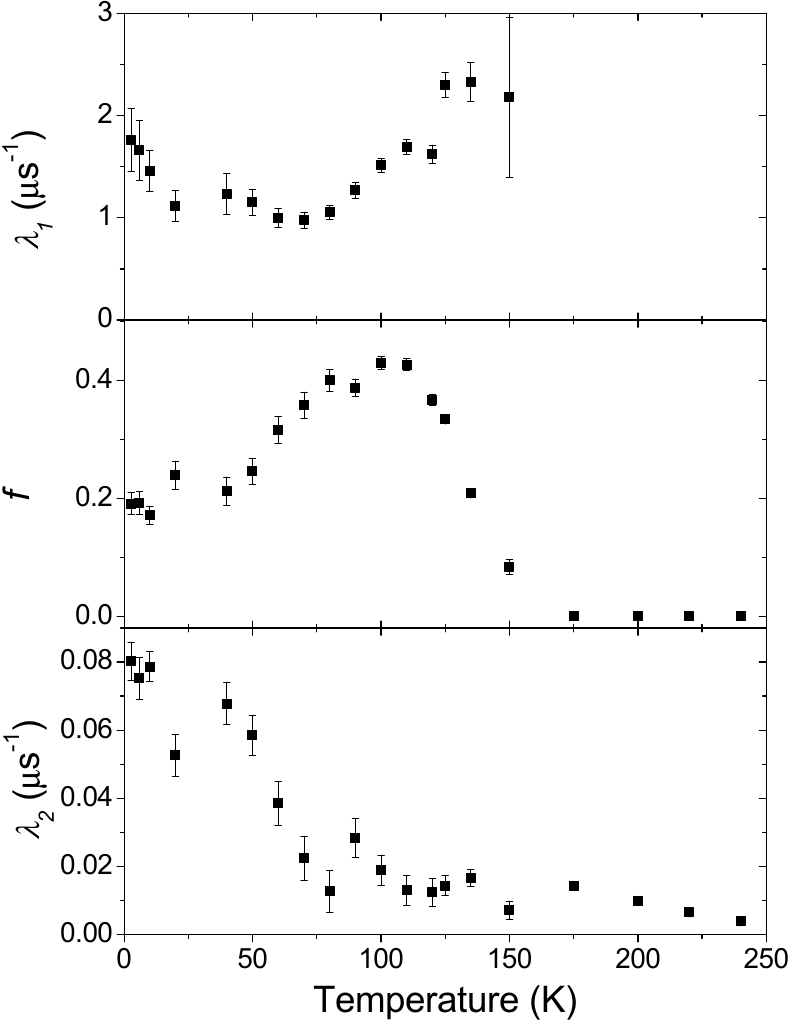}
\caption{The temperature dependence of $\lambda_1$, $\lambda_2$ and
  $f$ obtained from fits of the ZF spectra to Eq.~(\ref{fit_formula}).
  $f$ was set to zero for T$>$150~K.}
\label{Y2Ti2O7_fit_parameters}
\end{figure}

We base our analysis on a multi-state model and, for simplicity, we
restrict ourselves to a two-state model.\cite{Borghini78,Kehr78} In
this model the muon can be diffusing in the undisturbed regions or
trapped near a magnetic impurity. The relaxation rate from muons in
the trapped state is expected to be higher since they are closer to
magnetic defects and therefore experience stronger magnetic fields.
Also, muons in undisturbed regions can diffuse and reach trapping
sites (at a given trapping rate), while trapped muons can escape traps
if the temperature is high enough (at a given detrapping rate). The
muon polarization function for such a two-state model can be
approximated by the sum of two relaxing signals,\cite{Kehr78}
\begin{equation}
P_Z(t) = f \exp  \left (-\lambda_1 t  \right ) + 
(1 - f)  \exp \left (-\lambda_2 t  \right ), 
\label{fit_formula}
\end{equation}
where $\lambda_1$ and $\lambda_2$ are the relaxation rates of the two
components and $f$ is the contribution of the first component to the
full signal. At the limit of zero trapping/detrapping rates,
$\lambda_1$ and $\lambda_2$ represent the relaxation rate of muons in
traps and undisturbed regions, respectively, while $f$ is the fraction
of muons in traps. Eq.~(\ref{fit_formula}) provides a good qualitative
description of the ZF data at all temperatures as shown by the solid
lines in the top panel of Fig.~\ref{Y2Ti2O7_ZF_LF50_spectra}. The
temperature dependence of the three parameters in ZF is shown in
Fig.~\ref{Y2Ti2O7_fit_parameters}. We find $f \simeq 0.20$ and
temperature independent below 40~K, indicating that the muon trapping
and detrapping rates are constant below 40~K. Furthermore, the fact
that for these temperatures the signal can be decoupled by small
applied fields, is evidence that most of the muons cannot diffuse to
reach a high field trapping site during the experimental time window
(8.5$\mu$s). Therefore, we conclude that below 40~K the trapping and
detrapping rates must be small. At higher temperatures $f$ increases
due to the enhanced muon trapping caused by a faster diffusion in the undisturbed
regions, and then goes to zero at 135~K indicating that muons trap and
detrap so fast that they experience fast fluctuating magnetic fields
(motional narrowing) and relax following a single exponential
behaviour.\cite{Kehr78} We want to point out that fitting the data
with a temperature independent $f$ does not produce satisfactory fits;
and fitting with a power-exponential function produces very bad
results for temperatures above 70K.
Also, an implementation of the two state model as presented in
Ref.~[\onlinecite{Kehr78}] can not account for the observed
temperature dependence of $f$; probably because of the assumption that
the relaxation rate within each region (state) is exponential in the
whole temperature range.

To study the dynamic behavior at 2.4~K, we fit the LF data using the
polarization function (see Fig.~\ref{Y2Ti2O7_LF_spectra_decoupling}),
\begin{eqnarray}
P_Z(t) & = & f * P^{KT}(\Delta_1,\nu_1,B_{\rm ext},t) \nonumber \\
&& + (1-f) P^{KT}(\Delta_2,\nu_2,B_{\rm ext},t),
\label{e2}
\end{eqnarray}
where $P^{KT}$ is the analytical approximation of a Gaussian
Kubo-Toyabe function proposed by Keren.\cite{Keren94} By construction,
this equation assumes two independent muon fractions and therefore it
represents the two-state model in the limit of zero trapping and
detrapping rates. In each region (or state) the muons sense random
fields from a Gaussian distribution of width $\Delta_i$. This field
can fluctuate at a rate $\nu_i$ due to fluctuations of the magnetic
defects or hopping of the muon from one site to an other within the
same region. In the fit, $f$ was fixed to its ZF value (0.2) and
$B_{\rm ext}$ to the applied LF. The fit is good and completely
captures the decoupling of the signal, further supporting our
assumption of small muon trapping and detrapping rates. The values of
the fitted parameters are: $\Delta_1=3.00(5)$~mT,
$\nu_1=2.4(2)$~$\mu$s$^{-1}$, $\Delta_2=0.318(5)$~mT and
$\nu_2=1.52(7)$~$\mu$s$^{-1}$. $\Delta_1$, which correspond to the 
trapped state, is consistent with the
dipolar field expected in a region of 8~\AA\ around a 1$\mu_B$
magnetic impurity (see discussion below); and $\nu_2$ imposes a maximum
limit in the hopping rate of muons in the undisturbed regions.  We
want to note though that $\nu_1$ and $\nu_2$ are very similar and, in
fact, the data can be fit with a common fluctuations rate with no
significant effect on the quality of the fit. Since the contribution
from the paramagnetic defects to $\nu_1$ and $\nu_2$ is the same, this
indicates that the hopping rate in the undisturbed regions is probably
much smaller than $\nu_2$.

One point that needs to be discussed is the relatively large $f$
fraction found at low temperature. There are sixteen Y$^{3+}$ ions per
cubic unit cell of volume $V_{\rm cc} = a^3$. Let us denote $y$ the
percentage of magnetic defects relative to the Y$^{3+}$ population: $y
= x/2$, {\it i.e.} there are $16 y = 8 x$ Ti$^{3+}$ ions in $V_{\rm
  cc}$. Let us assume that a muon is trapped in a domain of relative
weight $f$ when implanted within a distance $d$ from a defect, and
also that the size and number of these domains are small enough such
that they do not overlap. Therefore, the volume around a defect in
which a muon is trapped is $V_\mu = 8 x (4 \pi/3) d^3$. Since $f =
V_\mu/V_{\rm cc}$, we conclude that $d = [(3/ 32 \pi) (f/x )]^{1/3}a$.
With $f \simeq 0.20$ and $y= 0.6 \%$ (see caption of
Fig.\ref{Y2Ti2O7_magnetization}), we compute $d = 0.79 a = 8$~\AA\
which is a reasonable value.\cite{Sanna04} This relatively long-range
influence of a defect is due to the long range of both, the dipolar
interaction and the nature of the elastic distortion
field.\cite{Eshelby56,Leibfried78}

In conclusion, we find that the muon hopping rate at 2.4~K, in both
stopping regions, is smaller than $\approx$2$\mu$s$^{-1}$. This value
is far below that observed for coherent diffusion in metals as well as
the $\sim$10$^3\mu$s$^{-1}$ proposed for coherent muon diffusion of
positive-muons in Dy$_2$Ti$_2$O$_7$.\cite{Quemerais12} Therefore we
find it unlikely that coherent muon diffusion is present in
Y$_2$Ti$_2$O$_7$. Nevertheless, an appropriate theoretical microscopic
calculation for the behaviour of positive-muons in this system is
needed to completely rule out or confirm coherent muon diffusion.
However, any theory should include the extended magnetic defects found
by our measurements. In this regard, our observations add to the
growing evidence that defects in pyrochlores are important to
understand their magnetic
properties\cite{Chapuis09a,Chapuis10,Yaouanc11a,Takatsu12,Yaouanc11c,Ross11a,Ross12},
and in particular those of Dy$_2$Ti$_2$O$_7$.\cite{Revell13} Finally,
an analysis to extract trapping and detrapping rates in the whole
temperature range would have to follow the lines of that in
Ref.~[\onlinecite{Kehr78}] but allowing for non-exponential relaxation
rates and/or going beyond the two states model.

We would like to thank 
Marisa Medarde for her support in the magnetization measurements and
A. Amato for the $\mu$SR assistance.

\bibliography{Y2Ti2O7_letter_b}

\end{document}